\documentclass[twocolumn,showpacs,footinbib,prl,floatfix]{revtex4}

\newcommand{\mm}{\mathrm}

\usepackage[pdftex]{graphicx}

\usepackage{sidecap}
\usepackage[center]{subfigure}

\usepackage{rotating}

\begin{document}

\title{Crystal plasticity as a mean field depinning transition: results from a phase field crystal model}

\author{Georgios Tsekenis, Thomas F. Fehm, Jonathan T. Uhl, Nigel Goldenfeld and Karin A. Dahmen}
\affiliation{Department of Physics, University of Illinois at
Urbana-Champaign, Loomis Laboratory of Physics, 1110 West Green
Street, Urbana, Illinois, 61801-3080.}

\begin{abstract}
Until now, most of our knowledge about the universality class
of crystal plasticity has come from simulations using discrete dislocation
dynamics.  These are force-controlled, typically at zero temperature, and deal
with the creation and annihilation of dislocations phenomenologically. In this
work, we go beyond these limitations by using phase field
crystal simulations in two dimensions at finite temperature to extract
the avalanche statistics of a simulated crystal under constant shear
velocity. In addition to the avalanche size and energy distributions
we extract the avalanche duration distributions and power spectra. 
All exponents and scaling functions extracted here for the statics 
and dynamics of crystal plasticity, belong to the mean field 
elastic depinning universality class, confirming earlier findings based on discrete 
dislocation dynamics.


\end{abstract}




\maketitle

Crystal plasticity is in some sense a solid state analogue of fluid
turbulence, with deformation at micro scales being both
intermittent and spatially inhomogeneous \cite{MiguelNat01, WeissMSEA01, RichetonNatMat05,
Richeton06, Richeton05, WeissJGR00, Weiss97, weiss2003sfa,
DimidukSci06, ZaiserAdvPhys06, TsekenisEPL13}.
These phenomena have been captured realistically by a number of
approaches, including discrete dislocation dynamics (DDD) models
\cite{MiguelNat01, ZaiserYielding, MiguelMSEA01, Laurson06,
CsikorSci07, Weygand2010, TsekenisJam11,TsekenisEPL13},
continuum models \cite{ZaiserYielding},
phase field models \cite{KoslowskiPRL04, KoslowskiPhilMag07}
and phase field crystal (PFC) models \cite{ChanPRL10}.
The most prominent open question that remains unanswered
is the one that unifies the accumulated literature
and solidifies the hard-earned knowledge: What is the
universality class of crystal plasticity?

A wealth of critical exponent and scaling function information from DDD simulations
(in two \cite{ZaiserYielding,TsekenisEPL13} and three dimensions \cite{CsikorSci07})
and experiments on slowly compressed nanocrystals and microcrystals
\cite{DimidukSci06, NirnanoPRL2012, ZaiserAdvPhys06}
strongly suggest that crystal plasticity belongs
to the mean field depinning universality class.
Nevertheless the issue is by no means settled;
for example recent DDD simulations in 2D obtained
non-mean field results \cite{IspanovityPRL14}.

Our main aim in this paper is to uncover the universal behavior
of deforming crystalline matter as it emanates from microscopic origins
and percolates through all scales. To this end we study crystalline plasticity
 with a phase field crystal model \cite{Elder04,Elder07,Chan09,ChanPRL10}.
The elementary entity in our simulations is a phase field representing
local atomic density, which is appropriately constrained to behave as
an atomic crystal, and indeed can be related to density functional
theory\cite{Elder07}. The phase field crystal exhibits elastic,
reversible deformation at small external loads. It deforms plastically
at exceedingly large external loads. Large deformations imprint
permanent, irreversible change in the lattice structure. At sufficient
shearing the periodicity is broken and topological defects emerge in
the system. Dislocations travel and interact with each other through
the lattice forming intricate structures. One can observe plastic
deformation being mediated through intermittent dislocation motion,
i.e. through discrete slip-avalanches. Here we extract several avalanche
measures and show that they are distributed according to power-laws
over several orders of magnitude revealing long range spatial and
temporal correlations. The set of critical exponents we calculate 
(the duration distribution power law exponent for the first time) 
fully supports the mean field depinning picture for
crystal plasticity. This is in strong agreement with earlier 2D DDD
simulations \cite{TsekenisEPL13}. It also agrees with 3 dimensional
simulations of dislocation dynamics \cite{CsikorSci07}, with
analytics \cite{DahmenUhlPRL09, ZaiserAdvPhys06} and experiments
\cite{NirnanoPRL2012, DimidukSci06, ZaiserAdvPhys06}. Our work
addresses the fundamental question of the universality class of crystal
plasticity and is relevant to the deformation of nano-crystals
\cite{GreerPRL08,GreerJenningsPRL10,NirnanoPRL2012} and micro-crystals
\cite{MiguelNat01, DimidukSci06, ZaiserAdvPhys06} as the need for
miniaturization of devices expands both in breath and depth.


\begingroup
\begin{table*}[t]
\centering
\begin{tabular}{cc|c|c|c|c|c}
\multicolumn{1}{c}{quantity}&\multicolumn{1}{c}{exponent}&\multicolumn{1}{c}{DDD sims}&\multicolumn{1}{c}{our PFC sims}&\multicolumn{1}{c}{MFT}& \multicolumn{1}{c}{other sims}&\multicolumn{1}{c}{experiments}\\
\hline \hline
$D_{S}(S) \sim S^{-\kappa}$   & $\kappa$      & 1.5 \cite{TsekenisEPL13} & 1.5 (Fig. \ref{fig:DofTSE_v}) & $\frac{3}{2}$ & 1.4\cite{ZaiserYielding},1.6\cite{MiguelMSEA01},1.5\cite{CsikorSci07}* & 1.5-1.6\cite{DimidukSci06},1.5\cite{GreerPRL08,NirnanoPRL2012}\\
$S_{\mm{max}} \sim \left ( 1-\frac{\tau}{\tau_{c}} \right )^{-\frac{1}{\sigma}}$   & $\frac{1}{\sigma}$  & 2 \cite{TsekenisEPL13} & & 2 & 2\cite{ZaiserYielding},2\cite{CsikorSci07}* & 2\cite{ZaiserYielding,NirnanoPRL2012} \\
\hline
$D(V_{\mm{max}}) \sim V_{\mm{max}}^{-\kappa_A}$  & $\kappa_A$ & & & 2\cite{LeBlancPRL12} & 1.8\cite{KoslowskiPRL04, KoslowskiPhilMag07} & 2\cite{WeissMSEA01}, $2.0\pm0.1$\cite{Richeton05}\\
"                                     & & & & " & & 1.5-2\cite{Weiss97},1.2-2.2\cite{WeissJGR00}\\
\hline 
$D_{t}\left( t_{\mm{aval}} \right) \sim t_{\mm{aval}} \hspace{0mm} ^{-1-\frac{\kappa-1}{\sigma \nu z}}$       & $\kappa_{t}=1+\frac{\kappa-1}{\sigma \nu z}$ & & 2 (Fig. \ref{fig:DofTSE_v}) & 2 & & \\
$t_{\mm{aval},\mm{max}} \sim \left ( 1-\frac{\tau}{\tau_{c}} \right ) ^{-\nu z}$          & $\nu z$                         & & & 1 & & \\
\hline 
$D_{E}(E) \sim E^{-\frac{1+\kappa-\sigma \nu z}{2-\sigma \nu z}}$     & $\kappa_{E}=1+\frac{\kappa-1}{2-\sigma \nu z}$ & 1.3 \cite{TsekenisEPL13} & 1.3 (Fig. \ref{fig:DofTSE_v}) & $\frac{4}{3}$ & $1.8 \pm .2$\cite{MiguelNat01}$^{+}$ &$1.6\pm0.05$\cite{MiguelNat01},$1.5\pm.1$\cite{Richeton06}\\
$E_{\mm{max}} \sim \left ( 1-\frac{\tau}{\tau_{c}} \right )^{-\frac{2-\sigma \nu z}{\sigma}}$    & $\frac{2-\sigma \nu z}{\sigma}$ & 3 \cite{TsekenisEPL13} & & 3 & & \\
\hline 
$\langle S \rangle \sim t_{\mm{aval}} \hspace{0mm} ^{\frac{1}{\sigma \nu z}}$    & $\frac{1}{\sigma \nu z} $ & $ 2 $ \cite{TsekenisEPL13} & 2 (Fig. \ref{fig:ps_svst_v}) & 2 & 1.5\cite{Laurson06}$^{+}$ & \\
$\langle t_{\mm{aval}} \rangle \sim S^{\sigma \nu z}$    & $\sigma \nu z$ & $0.5$ \cite{TsekenisEPL13} & & $\frac{1}{2}$ & & \\
\hline 
$V(t)_{\mm{shape}} \sim t_{\mm{aval}} \hspace{0mm} ^{\frac{1}{\sigma \nu z}-1}$   & $\frac{1}{\sigma \nu z}$ & $\sim 1.9 $ \cite{TsekenisEPL13} & & 2 & 1.5\cite{Laurson06}$^{+}$ &\\
$PS$,$PS_{\mm{int}}(\omega) \sim \omega^{-\frac{1}{\sigma \nu z}}$ & $\frac{1}{\sigma \nu z}$ & $2$ \cite{TsekenisEPL13} & 2 (Fig. \ref{fig:ps_svst_v}) &  2 & 1.5\cite{Laurson06}$^{+}$ & \\
\hline 
$\langle S^{m} \rangle \sim L^{\frac{1+m-\kappa-\sigma}{\nu \sigma}}$ & $\nu$ & $1.0\pm0.2$ \cite{TsekenisEPL13} & & 1 &  & \\
\hline
$\langle V \rangle \sim \left ( \frac{\tau}{\tau_c}-1 \right ) ^ {\beta}$  & $\beta$ & $1.1\pm0.1$ \cite{TsekenisEPL13} & & 1 & $1.8$\cite{MiguelPRL02}$^{+}$ & \\
\hline \hline
\end{tabular}
\caption{ Table of exponents. Our results from 2D PFC simulations are shown in the fourth column while our results from 2D DDD simulations are shown in the third column. Mean field interface depinning values are in the fifth column. Results from a full 3D DDD simulation are indicated with an asterisk (*). Results from a 2D DDD simulation with creation and annihilation in the steady state are indicated with a plus sign ($^{+}$). Symbol definitions: $D(x)$ stands for the distribution of $x$, $x_{\mm{max}}$ is the maximum of the distribution of $x$, $PS$ stands for power spectrum, $\left < x \right >$ stands for average of $x$, $S$ is the size of a slip avalanche, $t_{\mm{aval}}$ is its duration, $E$ its energy. $V(t)=\sum^{N}_{i=1}|v_{i}(t)|$ is the collective dislocation speed. $\tau$ stands for shear stress, $\tau_c$ is its critical value. Small greek letters are used for critical exponents throughout.\\
}
\label{table1}
\end{table*}
\endgroup

{\it The Phase Field Crystal Model, Sheared:}
The phase field crystal (PFC) model \cite{Elder02,Elder04,MAJA2007}
describes how the local density of atoms changes with time while
maintaining the symmetries and periodicity of the lattice. In addition,
the elastic interactions of the atoms are also captured by the PFC
allowing for the elasticity of the crystal to be expressed. These
characteristic properties of the phase field crystal model are
distinctly different from the phase field model. In a typical
phase field model the phase field describes the dynamics of interfaces
that separate dissimilar regions without keeping track of the
microscopic information inside those regions. In
\cite{KoslowskiPhilMag07, KoslowskiPRL04} Koslowski et al. developed a
phase field model to simulate dislocations as interfaces
(separating crystal regions with different accumulated slip). In
studying plasticity, however, it is important to capture the microscopic details
such as the dislocations which disrupt the periodicity of the perfect
lattice. At the same time, it is important to capture the macroscopic
behavior as well such as the collective motion of the dislocation
ensemble. The phase field crystal model is particularly successful in
doing that in an elegant way \cite{ChanPRL10}.

The total free energy in the phase field crystal (PFC) model
\cite{Elder02,Elder04}
\begin{equation}
F \{ \rho \}= \int{ \left [ \frac{\rho}{2}(\nabla ^2 + 1)^2\rho + \frac{r}{2}\rho^2 + \frac{\rho^4}{4} \right ] }d^dx
\label{pfcfreeenergy}
\end{equation}
is a functional of $\rho(\vec{x},t)$ the local density of the phase
field (at point $\vec{x}=(x,y)$ in space and time $t$). The first term
in Eq.(\ref{pfcfreeenergy}) penalizes departures of $\rho(\vec{x},t)$
from periodicity, thus describing a crystal structure as a density
wave. The last two terms impose a double well potential (to lowest
order) similar to the Landau ansatz.  The reduced temperature $r$ is
given by $(T-T_{c})/T_{c}$ and controls the phase behavior. The
material is liquid for temperatures higher than a critical temperature
$T_c$ while it crystallizes for temperatures below $T_c$.  Thus, for
$r>0$ one finds a liquid, constant $\rho$, phase (because the potential
is single well) while for $r<0$ the stable state is a triangular
lattice or a striped phase (due to the double well potential).

The PFC density $\rho(\vec{x},t)$ evolves according to the massive
phase field crystal equation
\cite{Stefa06,ChanPRL10}:
\begin{equation}
\frac{\partial^2 \rho}{\partial t^2} + (\beta)
\frac{\partial \rho}{\partial t}
 = (\alpha)^2\nabla ^ 2\frac{\delta F}{\delta \rho} + v(y)\frac{\partial \rho}{\partial x} + \eta,
\label{eqn_shear_PFC}
\end{equation}
where $\alpha$ controls the range and $\beta$ the time scale of
phonon excitations of the crystal \cite{Stefa06}. Thermal fluctuations
are represented by the stochastic noise $\eta$ which is assumed to be
Gaussian with second moment given by the fluctuation-dissipation theorem
$\langle \eta(\vec{x},t) \eta(\vec{x}',t') \rangle = -\epsilon \nabla ^2\delta(\vec{x}-\vec{x}')\delta(t-t')$.
The noise amplitude sets the scale of temperature $\epsilon \sim k_BT$.
This free energy is governed by conservative, relaxational, diffusive dynamics
that can be derived from density-functional theory \cite{MAJA2007}.

The PFC solid is a perfect triangular lattice in equilibrium, but its
excitations are phonons and topological defects such as dislocations.
When the phase field crystal is sheared, it will respond by generating
dislocations, just as a real crystal. The ability to create and
annihilate dislocations naturally and easily is one of the
advantages of the phase field crystal method, compared to DDD.

We applied a shear strain rate along the $x$ direction at the
$y=0,L_{y}$ boundaries by adding the convection term $v(y)\partial
\rho/\partial x$, to the evolution equation Eq.(\ref{eqn_shear_PFC}).
The boundary shear velocity profile,
$v(y)=\pm v_0 e^{\pm(y_0 - y)/L_y}$ ($y_0=0$ for $+$, $y_0=L_y$ for $-$)
is designed to be mainly controlled by the velocity at the boundary
$v_0$ since its penetration length $\lambda \ll L_y$ does not affect
the results strongly. The simulations take place in a square box of
sides $L_{x}$,$L_{y}$ in the $x$,$y$ direction. The boundary conditions
are periodic in $x$ and fixed at $y=0,L_{y}$ i.e. we design the
simulation cell such that the crystal wraps around in $x=0,L_{x}$
and terminates at $y=0,L_{y}$ (without wrapping around).
That way we can easily apply a fixed shear rate at the $y=0,L_{y}$ boundaries
and allow the dislocations to flow unbounded through the $x=0,L_{x}$ boundaries
effectively simulating a larger thermodynamic system than the mere dimensions of
our basic simulation cell.

The PFC model has the added value over DDD simulations
that it incorporates nonlinear elasticity \cite{chan2009nonlinear}
as well as dislocation creation and annihilation seamlessly without requiring 
additional phenomenological rules to model these number-changing operations.
The PFC methodology is thus uniquely capable of addressing such questions as 
how strain heterogeneity drives dislocation number fluctuations, 
which in turn couple to plasticity avalanches \cite{TarpPRL2015}.
Since the PFC model is in essence an atomic simulation, 
the highly nonlinear interaction at small dislocation distances 
is captured naturally through the crystal lattice that mediates it. 
The same holds true for the particular effects of creation
and annihilation of dislocations.

The PFC model handles applied shear velocity natively.
DDD models incorporate applied external stress naturally.
Thus the PFC is suitable to investigate the slip avalanches
above the critical point (flow stress) in the depinned state
while the DDD below the critical point, in the pinned state.
In that sense they perform complementarily to each other.

{\it PFC simulations at Finite Shearing Rate and Temperature:}
We study crystal plasticity as it proceeds intermittently through slip
avalanches using the sheared phase field crystal (PFC) model
\cite{ChanPRL10}.
We obtain the main scaling behavior of the distribution of a variety of
avalanche measures and for several different temperatures ($\epsilon$)
and shearing rates ($v_0$). We find remarkable agreement between
simulations and analytical mean field theory predictions of exponents
\cite{ZaiserAdvPhys06, FisherPR98, DahmenUhlPRL09}. Our results
strongly support the critical point picture of plasticity, and suggest
new experiments.

At every time step we obtain the phase field density $\rho$ in a 2D
square simulation cell (i.e. $L_{x}=L_{y}=L$) through
Eq.(\ref{eqn_shear_PFC}). Large values of $\rho$ indicates PFC \lq atoms" while low
field signifies interatomic space, cooperatively arranged into a tight
crystal (triangular in 2D). By applying shear along the fixed
boundaries $y=0,L$ the translational symmetry breaks and dislocations
are created in an attempt to relieve the high stress accumulated near
those boundaries. Once the dislocations are created, they interact with
each other to form pairs and more complex structures such as low-angle grain
boundaries; individual dislocations, dislocation pairs and grain boundaries can be
seen in the snapshot of the PFC simulations in Supplementary Material. 
Dislocations may also glide throughout the entire crystal, 
allowing for slips to self-organize into slip avalanches. 
We quantify the avalanche activity by extracting
the total speed of the dislocations,
\begin{equation}
V(t)=\sum_{i=1}^{N(t)} |\vec{v}_i|,
\end{equation}
where $N(t)$ is the number of dislocations in the system at time $t$
and $\vec{v}_i$ is the speed of the $i$-th dislocation. This measure
is similar to the acoustic emission signal in Weiss \textit{et al.}'s
single crystal ice experiments (e.g. \cite{MiguelNat01}).
Other variants of the avalanche activity measure in the literature
can track the avalanches as well. For example the collective dislocation
velocity is defined as $V'(t)=\sum_{i=1}^{N(t)} b_i \vec{v}_i$
($b_i$ is the dislocation's burgers vector) and represents
the strain rate of the crystal \cite{MiguelPRL02, ZaiserYielding, ZaiserAdvPhys06}.
Note that all dislocations, single dislocations or disclination pairs that move at high speeds
and grain boundaries that move slowly, participate equally in the calculation of $V(t)$ (see figure in Supplementary Material).


A perfect unsheared triangular crystal is devoid of dislocations and
therefore every atom has $n_i=6$ nearest neighbors (identified with
Delaunay triangulation). Conversely there exists one dislocation for
every atom with $n_i=5$ or $n_i=7$  neighbors since vacancies are not
allowed.

We capture the speed of the dislocations, $V(t)$, through the speed of these defect atoms with $n_i\neq 6$ neighbors,
$\tilde{V}(t) = \sum_{i=1}^{N_{\mm{defect}}(t)} |\vec{u}_i|$,
where $N_{\mm{defect}}(t)$ is the number of defect atoms and $\vec{u}_i$ is the velocity of defect atom $i$.

In order to partition the signal into individual slip avalanches we
apply a threshold, $V_{\mm{thr}}$, to it for each temperature and
shearing rate we simulate. The beginning of the avalanche is signified
at an instant when the collective dislocation speed $V(t)$ intersects upward
the threshold while its end is when $V(t)$ crosses the threshold downward
immediately after. We extract the probability distribution of the avalanche duration
\begin{equation}
t_{\mm{aval}} = t_{\mm{finish}} - t_{\mm{start}},
\end{equation}
size (also called activity fluctuations in the flowing state)
$S = \int_{t_{\mm{start}}}^{t_{\mm{finish}}} V(t) dt$,
and energy
$E = \int_{t_{\mm{start}}}^{t_{\mm{finish}}} V^2(t) dt$,
where $t_{\mm{start}}$ and $t_{\mm{finish}}$ are the starting and
ending time of the event respectively. In order to
see the fluctuations that correspond to slip avalanches we applied a
threshold equal to the average of the signal, $V(t)$, in each
realization of total time $t_{\mm{total}}$,
$V_{\mm{threshold}}=\frac{1}{t_{\mm{total}}}\int_{0}^{t_{\mm{total}}} V(t) dt$,
We also calculate the power spectrum of $V(t)$:
\begin{equation}
PS(\omega)=\left| \int_{0}^{t_{\mm{total}}} V(t) e^{-i \omega t} dt \right |^2.
\end{equation}
The power spectrum reveals the frequency content of the time series of
the collective dislocation speed and it needs no thresholding. It is
equivalent to extracting the time-time correlations of the collective
dislocation behavior and near a critical point it is expected to have
no characteristic scale, i.e. be scale-free \cite{Kuntz00,Travesset02}.

\begin{figure}[ht]
\includegraphics[width=0.95\columnwidth]{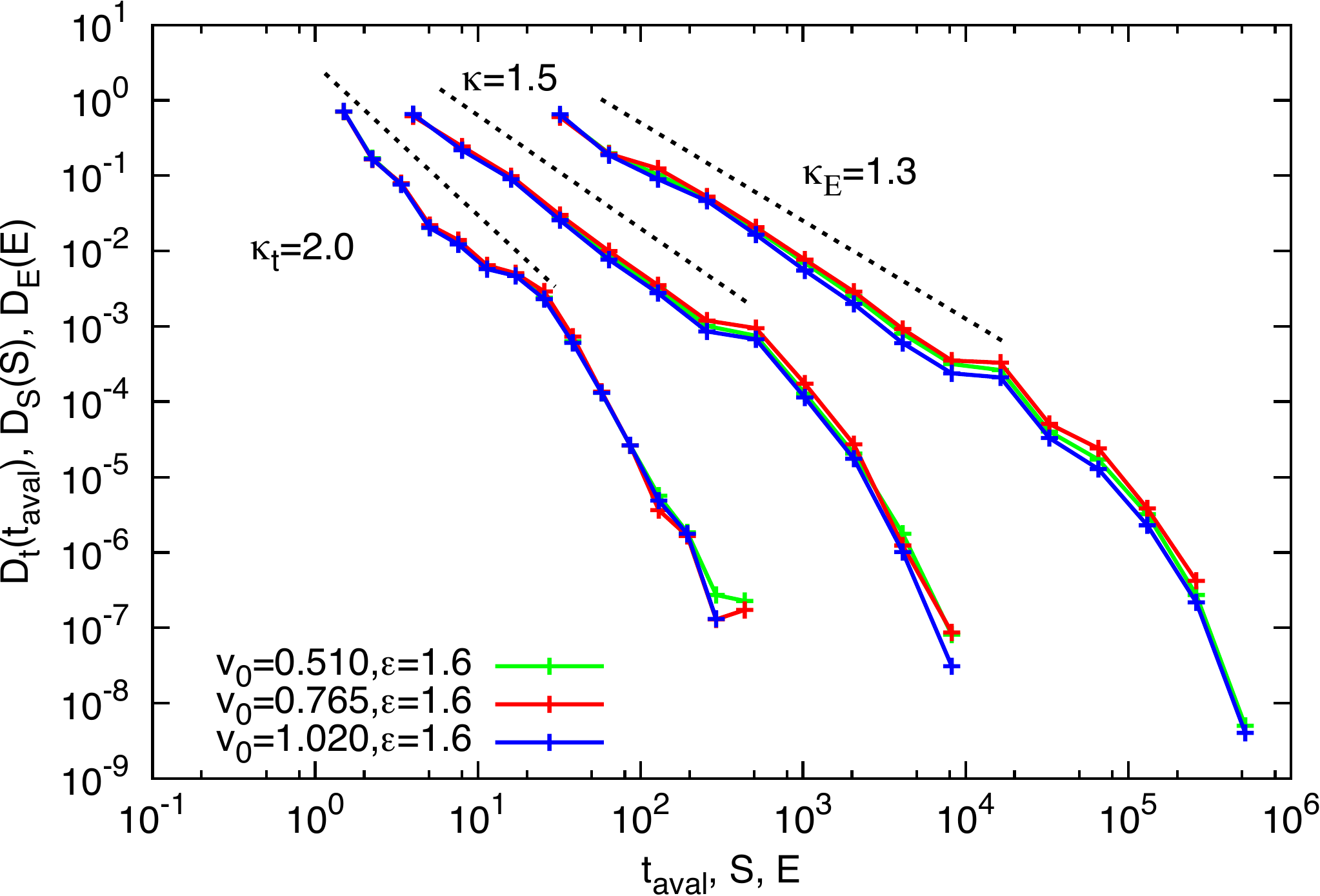}
\caption{ (color online) Probability distributions $D_{t}(t_{\mm{aval}})$ (left), 
$D_{S}(S)$ (middle) and $D_{E}(E)$ (right) of avalanche duration $t_{\mm{aval}}$,
size $S$ and energy $E$ respectively  for different shearing rates $v_0$ 
at the same temperature $\epsilon=1.6$. 
The probability distribution of the slip avalanche sizes
follows a power law with exponent $\kappa \approx 1.5$, of durations with $\kappa_{t} \approx 2$ 
and of energies with $\kappa_{E} \approx 1.3$.
These results are in agreement with MFT (Table \ref{table1}).
} \label{fig:DofTSE_v}
\end{figure}

\begin{figure}[ht]
\begin{center}
\includegraphics[width=0.91\columnwidth]{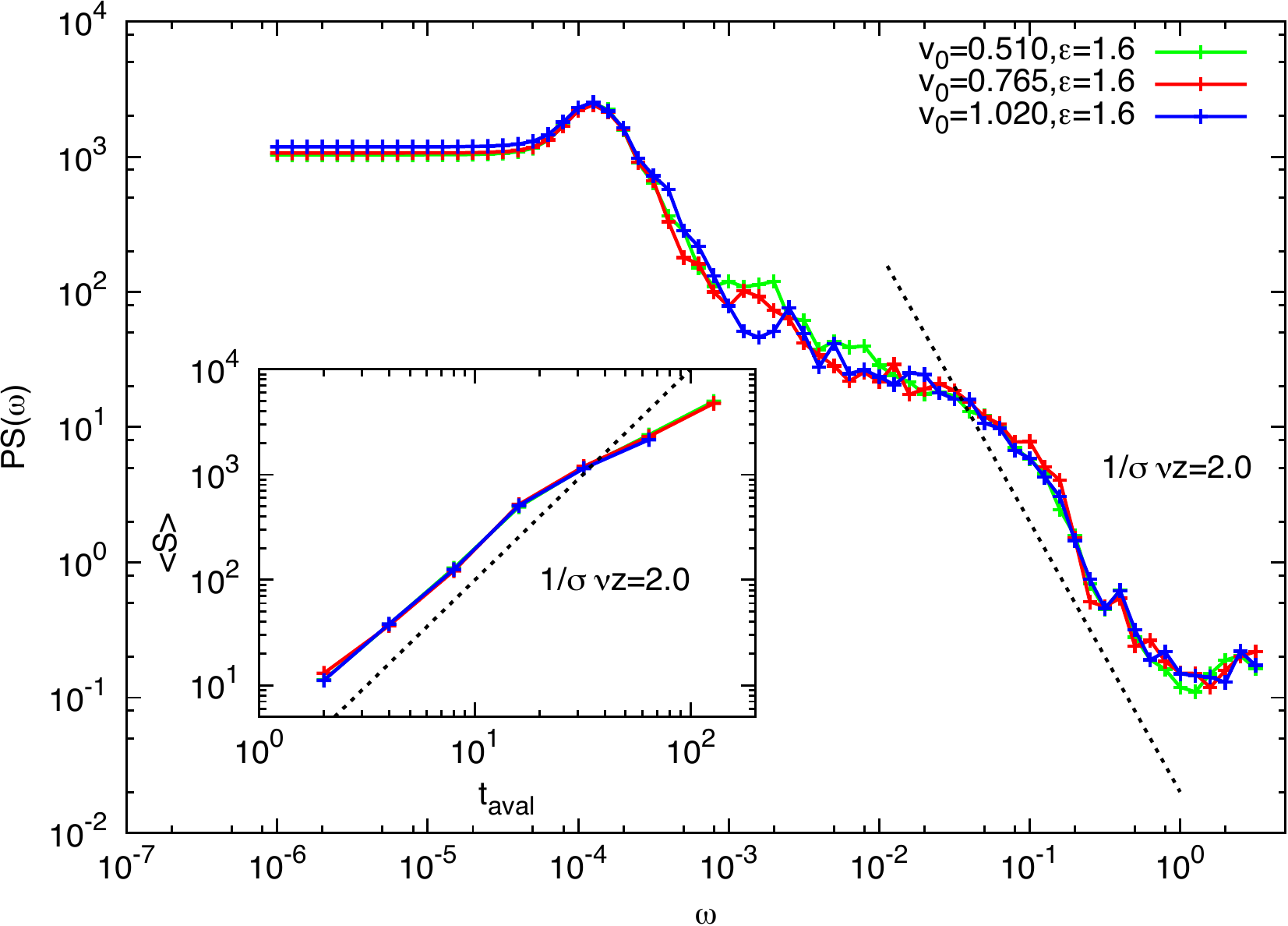}
\caption{ (color online) The power spectrum $PS(\omega)$ of the collective dislocation speed (main)
and the average slip avalanche size $\langle S \rangle$ versus duration $t_{\mm{aval}}$ (inset) 
for different shearing rates $v_0$ at the same temperature $\epsilon=1.6$. 
The power spectrum scales with the inverse square frequency giving $1/\sigma \nu z \approx 2$.
The average size scales with the square of the duration, $1/\sigma \nu z \approx 2$,
for avalanches that are sufficiently small not to touch the sample boundaries.
(The boundaries of the power-law scaling regime of the power spectrum $PS(\omega)$ 
are inversely proportional to the boundaries of the power-law scaling regime of the duration distribution $D_{t}(t_{\mm{aval}})$).
Results agree with MFT predictions (Table \ref{table1}).
} \label{fig:ps_svst_v}
\end{center}
\end{figure}

For each shearing rate and temperature, we run $48$ different realizations
each with different seed for the random number generator for the noise $\eta$
to obtain sufficient statistics. This results in tens of thousands of avalanche
events for each shearing rate $v_0$ and temperature $\epsilon$. 
In Fig. \ref{fig:DofTSE_v}  we show the event size, duration and energy distributions
while in Fig. \ref{fig:ps_svst_v} the power spectra and average size versus duration 
for different shearing rates at the same temperature. 
We find that the distributions follow a power law for small event sizes and cut off at larger sizes,
with the maximum avalanche size not exhibiting a strong dependence on
shear rate over the simulated range of the parameter $v_0$.
We suspect that the reason that shear rate does not affect
the distribution much is that for our systems the system size
sets the cutoff of the avalanche size distribution. For much larger
systems the mean field theory predicts that an increase in shear rate
will reduce the cutoff of the avalanche size distribution. This can
only be seen in systems that are so large that the system size is much
larger than the correlation length of the avalanches given by the
finite shear rate. In previous work with PFC \cite{ChanPRL10}
a different threshold was applied to the signal so as to extract avalanches.
It resulted in a rate-dependent avalanche distribution cutoff. Although both
thresholds reveal the same power-law in the avalanche distributions
we believe the threshold used here (equal to the average of the dislocation activity for each run)
is a more natural way to quantify the fluctuations around the mean dislocation activity.

{\it Scaling Behavior of the avalanches:}
The distributions of the avalanche size, duration and
energy, the power spectra and average size versus duration
shown at same shearing rate, $v_0=0.765$, 
and different temperature parameter values $\epsilon$ 
are shown in figures in the Supplementary Material. 
In Figs. \ref{fig:DofTSE_v} and \ref{fig:ps_svst_v} we presented the
distributions of the avalanche size, duration, energy, power spectra and average size
versus duration at the same temperature parameter values
$\epsilon=1.6$ and different shearing rates, $v_0$. Each curve is
characterized by a power law for several decades and a cutoff at large
values (smaller values for the power spectra) which does not change with shear rate.
Mean field theory predicts the dependence on the shear rate, that should be visible in
larger simulations \cite{KarinNatPhys09}.
The slip events distribute themselves according to power laws
$D_{t}(t_{\mm{aval}}) \sim t_{\mm{aval}}^{-2}$,
$D_{S}(S) \sim S^{-1.5}$,
$D_{E}(E) \sim E^{-1.3}$,
$PS(\omega) \sim \omega ^{-2}$,
with critical exponents that are in agreement with the Mean Field
interface depinning transition universality class (Table \ref{table1}).

Our extended results from the PFC model agree with the majority of the
robust experimental and computational results in the literature (for an
extended summary see Table \ref{table1}).
Friedman \textit{et al.} \cite{NirnanoPRL2012} analyzes the slip statistics 
of compressed crystalline nano-pillars in different stress bins 
as the flow stress is approached and get $\kappa=1.5$ (also $\sigma=2$).
Dimiduk \textit{et al.} measure $\kappa=1.5-1.6$ from compression experiments 
on micro-pillars at slowly increasing stress \cite{DimidukSci06}. 
Similarly, in 2D DDD and continuum models with quasi-static stress increase in the pinned regime,
Zaiser \textit{et al.} calculate $\kappa=1.4$ \cite{ZaiserYielding,Zaiser3}.
Slip-event energy amplitudes are power law distributed
with exponent $\kappa_{E}=1.8$ \cite{MiguelNat01} in simulations,
and $\kappa_{E}=1.6$ in experiments \cite{MiguelNat01}.
Richeton \textit{et al.} \cite{Richeton06} report $\kappa_{E}=1.5$ for
the energy distribution of acoustic emission deformation experiments.
Of course in \cite{TsekenisEPL13} a large number of scaling exponents
was extracted from 2D DDD simulations and as shown in Table
\ref{table1} they corroborate that crystalline plasticity 
belongs to the universality class of mean field depinning 
for the discussed static and dynamic properties of the avalanche statistics.
This result is also consistent with analytic results \cite{DahmenUhlPRL09,ZaiserAdvPhys06} 
and simulations in 3 dimensions \cite{CsikorSci07}.

{\it Discussion:} We approach crystalline plasticity with a new and
alternative simulation methodology: the phase field crystal model.
The PFC simulations can essentially be thought of
as molecular dynamics simulations but greatly sped up. At the same time
they are free from the several phenomenological rules
and constraints that discrete dislocation dynamics need in order
to incorporate the variety of dynamical phenomena that take place
in a stressed crystal. The reason is that the phase field crystal
reproduces faithfully the real crystal including its elastic properties
and topological behavior.

By employing this sheared PFC model we were successful in
extracting the characteristic statistical scaling behavior of plastic deformation.
We verified the robustness of the DDD simulations, nullified potential artifacts 
of the add-on phenomenological creation and annihilation rules 
of the DDD simulations and strengthened the universal conclusions.
Our results reaffirm that all the exponents and scaling forms extracted here 
for crystal plasticity are consistent with the mean field interface depinning 
universality class -- even in the absence of frozen-in pinning centers.

An increasing number of studies (including this work) have indicated the
striking similarities between crystal plasticity and the interface
depinning dynamic phase transition \cite{ZaiserAdvPhys06,
DahmenUhlPRL09, TsekenisEPL13}.
The critical exponents we found here are in
excellent agreement with the mean field theory of the interface
depinning universality class (see Table \ref{table1}) even though the
PFC results are performed at finite temperature, $\epsilon \sim k_{B}T$,
and as a result the extracted scaling relations are plagued by
larger fluctuations (due to temperature-induced dislocation creep),
when compared to DDD simulations at $T=0$.
{\it Acknowledgements:} We thank Pak Yuen Chan, Jonathan Dantzig and Stefano Zapperi
for helpful conversations. We acknowledge NSF grant DMR 03-25939 ITR (MCC) and DMR
10-05209, the University of Illinois Taub cluster.

\bibliographystyle{apsrev}

\bibliography{one}

\end{document}